\documentclass[final,5p,times,twocolumn]{elsarticle}

\usepackage{amsmath}     
\usepackage{amssymb}
\usepackage[dvipsnames]{xcolor}

\usepackage[figuresright]{rotating}
\usepackage{graphicx}

\usepackage{subfigure}

\begin{document}

\begin{frontmatter}






\title{Study of space charge phenomena in GEM-based detectors}


\author[First]{Promita Roy\fnref{label1}}
\ead{promita.roy08@gmail.com}

\author[Second]{Prasant Kumar Rout}

\author[Third]{Jaydeep Datta}

\author[Fourth]{Purba Bhattacharya}
\author[First]{Supratik Mukhopadhyay}
\author[First]{Nayana Majumdar}
\author[First]{and Sandip Sarkar\corref{cor2}}

\fntext[label1]{Corresponding author}
\cortext[cor2]{Retired Professor}

\address[First]{Saha Institute of Nuclear Physics, Kolkata-700064, a CI institute of Homi Bhabha National Institute, Mumbai-400094}
\address[Second]{National Central University, Taoyuan City 320317, Taiwan}
\address[Third]{Université Libre de Bruxelles, Av. Franklin Roosevelt 50, Belgium}
\address[Fourth]{School of Basic and Applied Sciences, Adamas University, Kolkata-700126}

\begin{abstract}
  {Space charge accumulation within GEM holes is one of the vital
phenomena which affects many of the key working parameters of the detector. This accumulation is found
to be significantly affected by the initial primary charge configurations and applied GEM voltages since they determine charge sharing and the subsequent evolution of detector
response. In this work,
we have studied the effects of space charge phenomena on different parameters for single
GEM detectors using a hybrid numerical model.  }
\end{abstract}

\begin{keyword}
Gas Electron Multiplier \sep simulation \sep charge sharing \sep space charge effect



\end{keyword}

\end{frontmatter}


\section{Introduction}
Gas Electron Multipliers (GEMs)~\cite{a} are being widely used in various high energy particle physics
experiments world-wide. The understanding of the functioning of
these GEM detectors are, thus, of general interest. Space charge accumulation within GEM holes is
one of the crucial factors apart from the charging up phenomena~\cite{b} which affects many of the key parameters of such detectors
through its direct influence on the resulting electric field in and around the holes. A recent numerical study~\cite{c} based on a 2D-axisymmetric model has indicated the possible effect of charge sharing on space charge accumulation in GEM holes. It was noted that charge sharing among a larger number of holes results in higher gain since the same amount of charge gets shared among many holes leading to less perturbation in the field. Moreover, it has been observed from~\cite{c} that ion driven space charge effect leads to
discharges in GEM detectors. This motivated us to study the space charge accumulation and it’s relation with the charge sharing phenomena thoroughly. 

\section{Simulation details}
A 2D-axisymmetric geometry of a GEM detector irradiated with a $^{55}$Fe source has been simulated with Argon-Carbon dioxide based gas mixture in the volumetric ratio of 70:30. Pressure and temperature equal to 1 atm and 298 K respectively have been used for the simulation. HEED~\cite{d} and GEANT4~\cite{e} have been used to identify the primary electron clusters and their positions in the detector volume whereas MAGBOLTZ~\cite{f} has been used to get the relevant transport properties for the chosen gas mixture. A seed cluster has been constructed as a 3D gaussian distribution using the information about the number of primary electrons and their positions obtained from the $^{55}$Fe radiation source. Space charge phenomena is influenced by the charge sharing effect within the GEM holes and the applied field across the GEM foil. The former has been calculated using GARFIELD~\cite{g} and the latter which is the calculation of electric field and the simulation of transport of electrons and ions within the detector volume have been done using a fast hydrodynamic numerical model~\cite{h} developed on the COMSOL Multiphysics platform\cite{i}. Electrons and ions are treated as charged fluids and space charge effect is incorporated in the electron/ion transport model of COMSOL. As a result it is capable of simulating variation in applied electric field due to the presence of electrons and ions within the hole.

\begin{figure}[htbp]
	\centering
	\includegraphics[width=8.5cm]{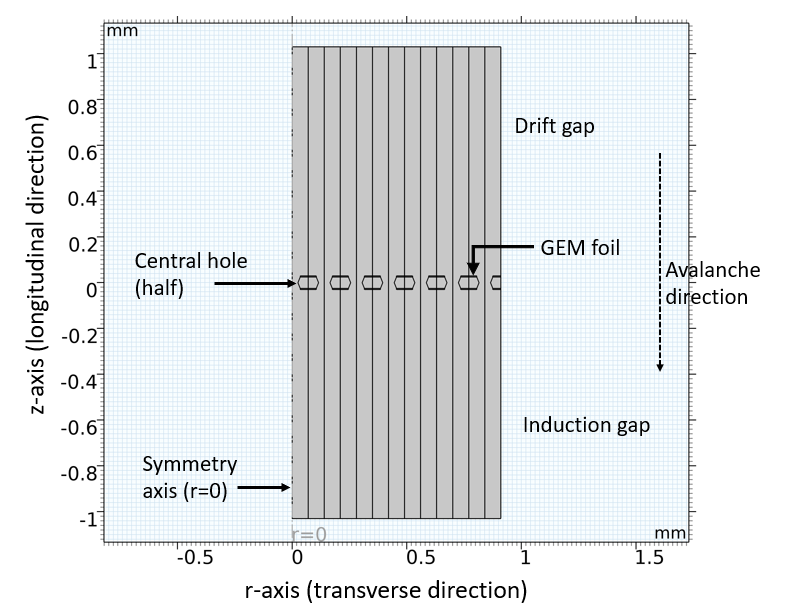}
	\caption{2D-axisymmetric geometry of a GEM detector}\label{fig1}
\end{figure}

The simulated GEM foil is a 50$\mu m$ thick kapton foil with 5$\mu m$ thick copper coating on both sides. It has biconical holes with diameters 70-50-70$\mu m $ and pitch of 140$\mu m$. Drift gap and induction gap have been chosen to be 1mm each. Figure~\ref{fig1} shows the simulated 2D-axisymmetric geometry of the GEM which has a half central hole and all other holes are channels symmetric about the central hole symmetry axis (r=0). Presence of these off-center channels instead of actual holes affects the electric field. The central hole being an actual hole has the correct field value, however, the off-center holes have field values less than the field of the central hole. The maximum deviation in the field is around 20\% of the actual field. To scale up the electric field and the charge distribution, scaling factors have been calculated using results from GARFIELD. This portion can be found in detail in section 3.2 and appendix A of the paper~\cite{h}.The resulting avalanche in this 2D-axisymmetric geometry is symmetric about the axis of the central hole.

The space charge effect was assessed from the point of view of:
\begin{itemize}
    \item Variation of spatial distribution and mean z-position of primary seed cluster in the drift region.
    \item Operating voltages of the detector.
\end{itemize}

 To take into account the spatial distribution of the primary seed clusters, five cluster cases have been chosen by considering the mean spread of primary seed clusters and corresponding sigma values in radial(r) and vertical(z) direction as given in figure 4 in~\cite{h}. The mean spread of the primary cluster was obtained to
be  $132 \mu m$ in radial direction and $154 \mu m $ along z direction. This mean primary cluster has been taken to be case 1 and all the other cases have been obtained by adding or subtracting $1\sigma$ of the fitted mean cluster in radial and z direction from their respective mean values.
The five different cluster configurations are modeled as follows:
 \begin{itemize}
    \item case 1: $\Delta$r = 132 $\mu$m,  
    $\Delta$z = 154 $\mu$m
    \item case 2: $\Delta$r = 203 $\mu$m,  
    $\Delta$z = 234 $\mu$m
    \item case 3: $\Delta$r = 61 $\mu$m,  
    $\Delta$z = 74 $\mu$m
    \item case 4: $\Delta$r = 203 $\mu$m,  
    $\Delta$z = 74 $\mu$m
    \item case 5: $\Delta$r = 61 $\mu$m,  
    $\Delta$z = 234 $\mu$m
\end{itemize}

At the start of the simulation each of these primary elecron clusters are released from a height of $250 \mu m$ above the GEM foil in the drift gap.

\section{Results}
\begin{figure}[htbp]
\centering
 \includegraphics[width=7cm]{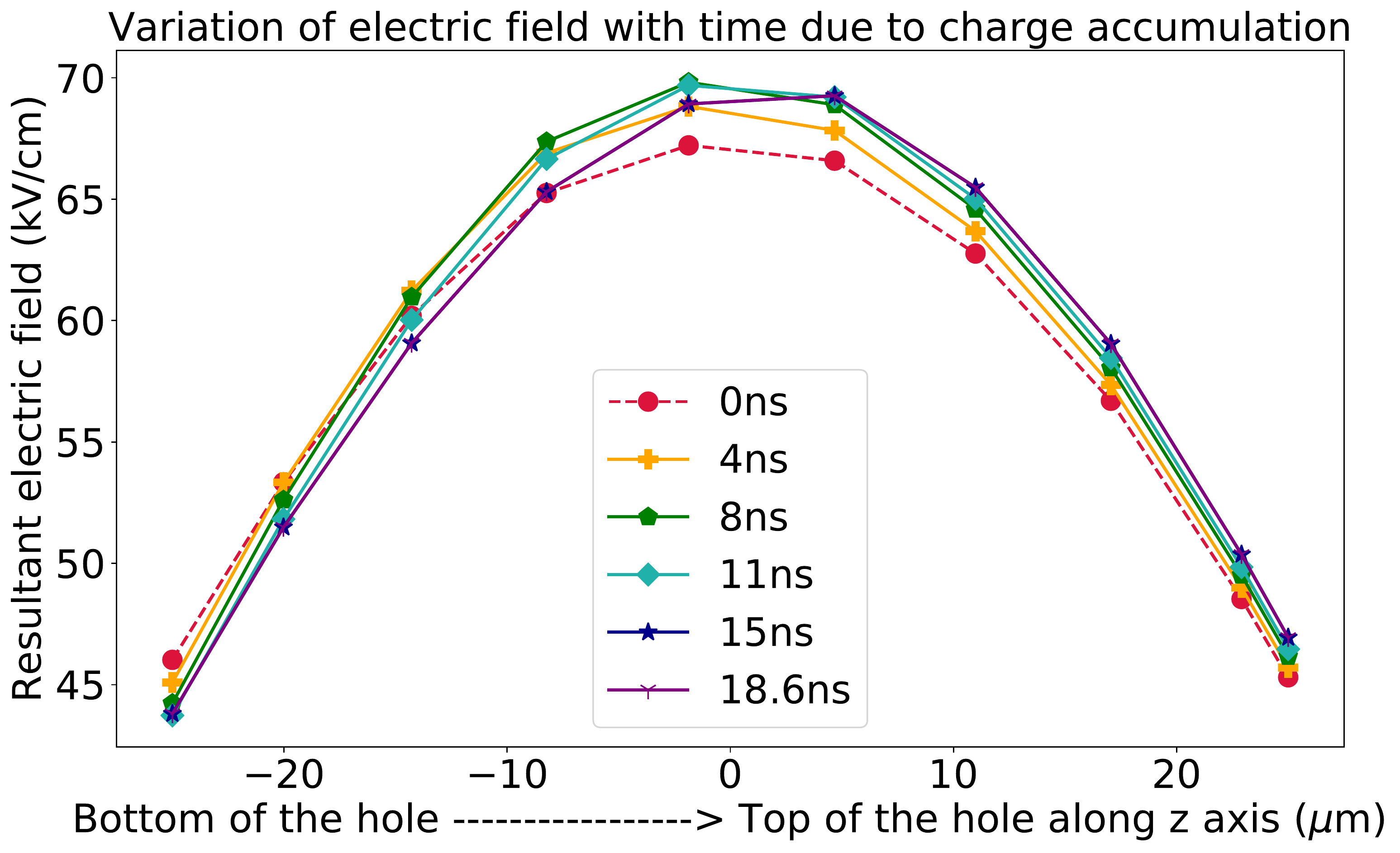}
  \caption{Variation of electric field due to space charge effect}\label{fig2}
\end{figure}

Accumulation of space charges inside the holes results in the modification of the applied electric field within the GEM foil which in turn can modify the effective gain~\cite{j}. Figure~\ref{fig2} shows the variation of electric field due to space charge effect inside the GEM holes at various time-steps. At "t=0" the primary electron cluster is just released from a height($250 \mu m$) in the drift, so the electric field at "t=0" has no effect of space charges. In the presence of applied electric field, the primary electrons drift and diffuse towards the GEM, enter the holes and get multiplied producing more number of electrons and ions. As a result the applied electric field inside the holes get perturbed as shown at different time stamps in the figure~\ref{fig2}. Due to this modification of the electric field inside the GEM holes, the value of effective gain of the detector gets modified. 

\begin{figure}[htbp]
\centering
 \includegraphics[width=8.5cm]{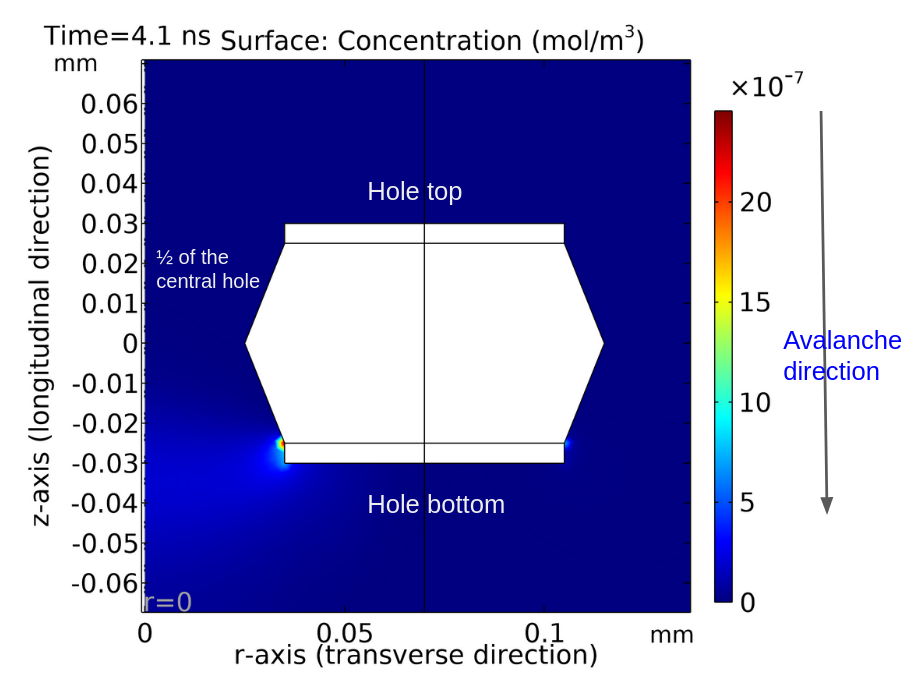}
  \caption{Position of avalanche electrons when the number of electrons is maximum}\label{fig3}
\end{figure}

Figure~\ref{fig3} shows the position of the avalanche electron cloud inside the holes at t = 4.1 ns when the electron multiplication is maximum. To see the variation of electric field due to space charges, the difference in electric field has been calculated as given below
\begin{equation}
	\centering
	\Delta E = E_w - E_{wo}  
\end{equation}\label{e1}

Here, $E_w$ is electric field with space charge effect and  $E_{wo}$ is electric field without space charge effect.

\begin{figure}
    \centering
    \includegraphics[width=7.3cm]{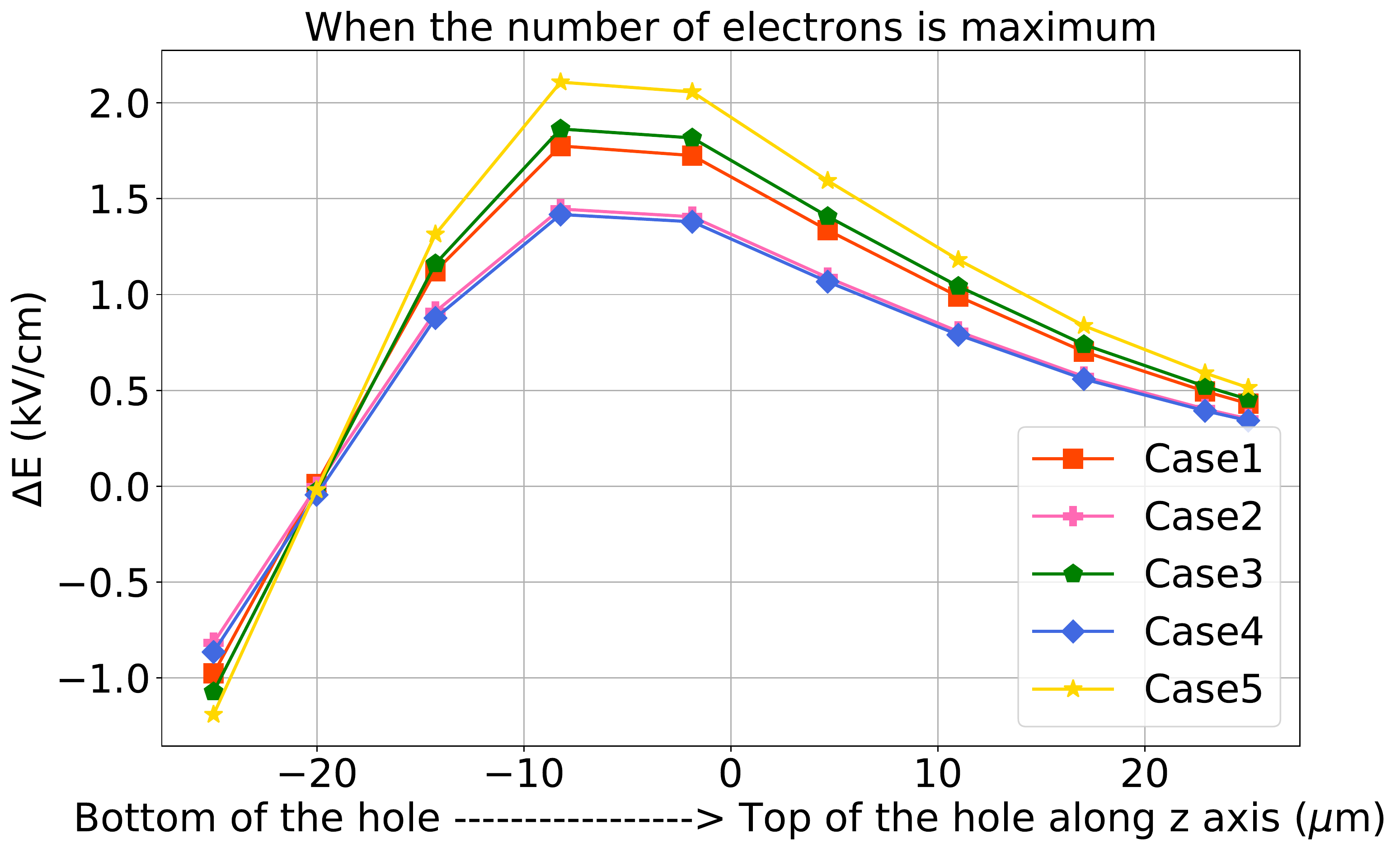}
    \includegraphics[width=7cm]{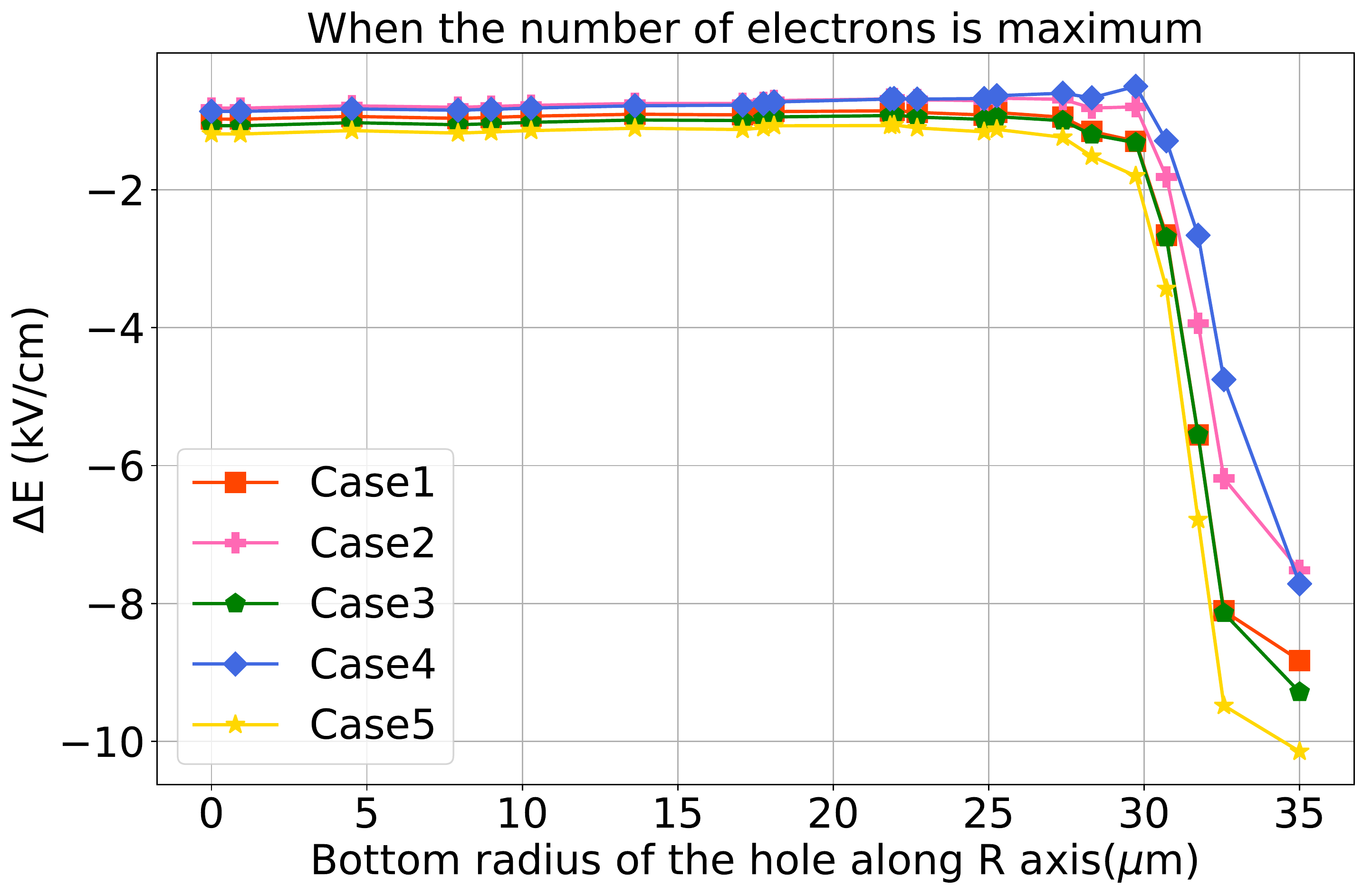}
    \caption{Variation of electric field along the height of the hole (\emph{top figure}) and  along the bottom radius of the hole (\emph{bottom figure}) due to space charge effect for different spatial distributions of the primary cluster released from a height of 250$\mu$m above the GEM foil}
    \label{fig4}
\end{figure}

Figure~\ref{fig4} shows the variation of electric field due to space charge effect for different primary seed cluster cases placed at a height of 250$\mu$m above the GEM foil. It has been observed that the difference in electric field ($\Delta E$) is more negative for cases which have less radial spread (case 3 and case 5) , whereas cases with larger radial spread (case 2 and case 4) have lower negative values of $\Delta E$. Thus, it is observed that case 3 and case 5 have lower resultant electric field values along the bottom of the hole. Since it is known from figure~\ref{fig3} that most of the electron multiplication inside the GEM occurs in the vicinity of the hole bottom, cases 3 and 5 end up having lower gain values and the other two cases which are radially extended (case 2 and case 4) have the higher gain values. This can also be interpreted as the cases which exhibit more charge sharing to end up having more gain and less space charge effect.

\begin{figure}[htbp]
	\centering
	\includegraphics[width=7.5cm]{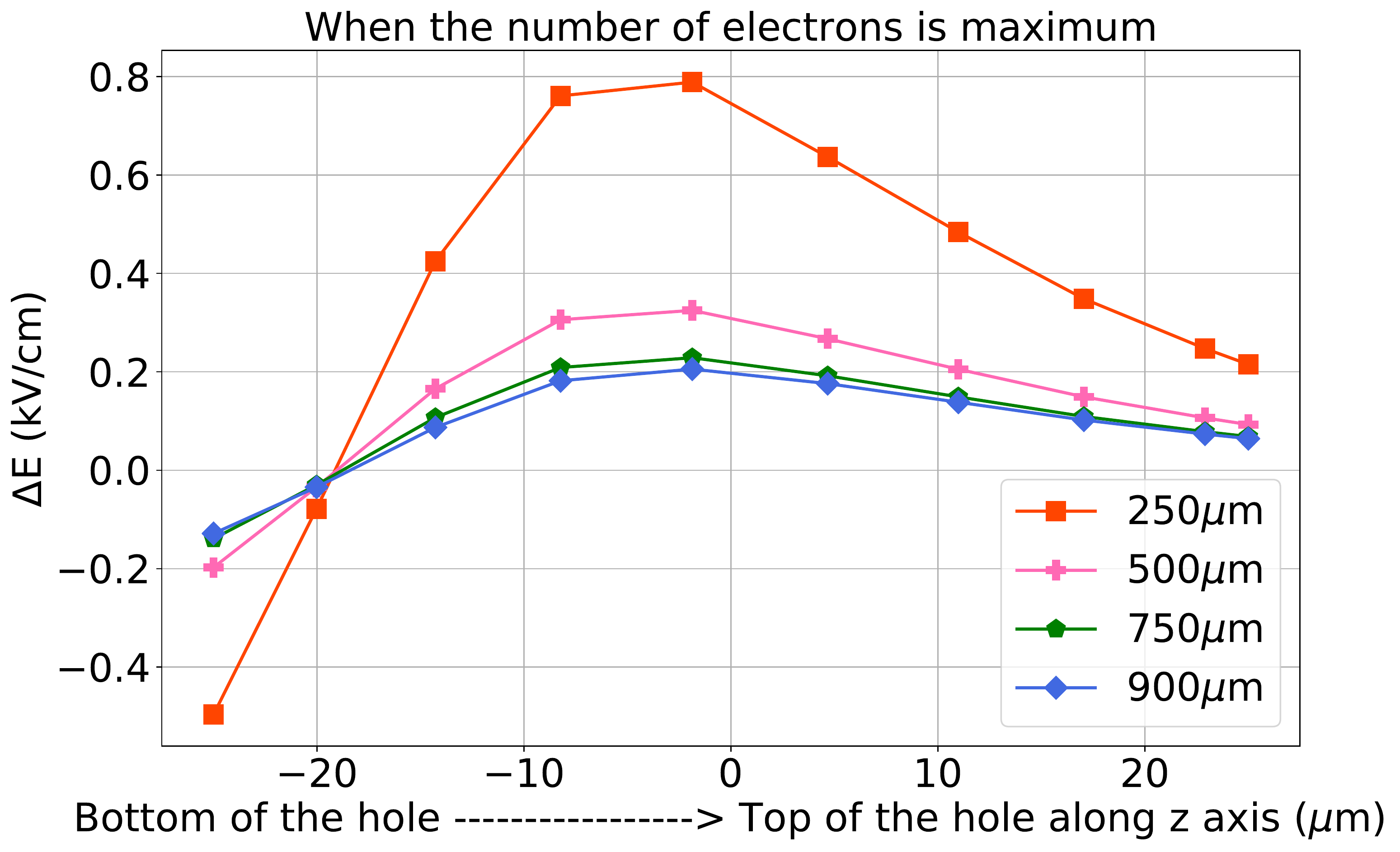}
	\caption{Variation of electric field due to space charge effect for different heights of the primary cluster in the drift gap for cluster case 1.}\label{fig5}
\end{figure}

Similarly, the Figure~\ref{fig5} shows the variation of applied electric field due to space charge effect for primary electrons clusters placed at different heights in the drift gap. Different heights correspond to different mean z-positions of the electron clusters above the GEM foil in the drift gap. Four such mean z-positions $250 \mu m$, $500 \mu m$, $750 \mu m$ and $900 \mu m$ in the drift gap have been chosen. It has been observed that the space charge effect decreases with increasing height of the primary cluster. When the primary cluster is closer to the GEM foil, diffusion as well as charge sharing is comparatively less and as a consequence space charge effect is more prominent which results in the decrease of the electric field around the bottom of the hole. Thus a primary seed cluster closer to the GEM foil ends up having comparatively less effective gain.

\begin{figure}[htbp]
	\centering
	\includegraphics[width=7.5cm]{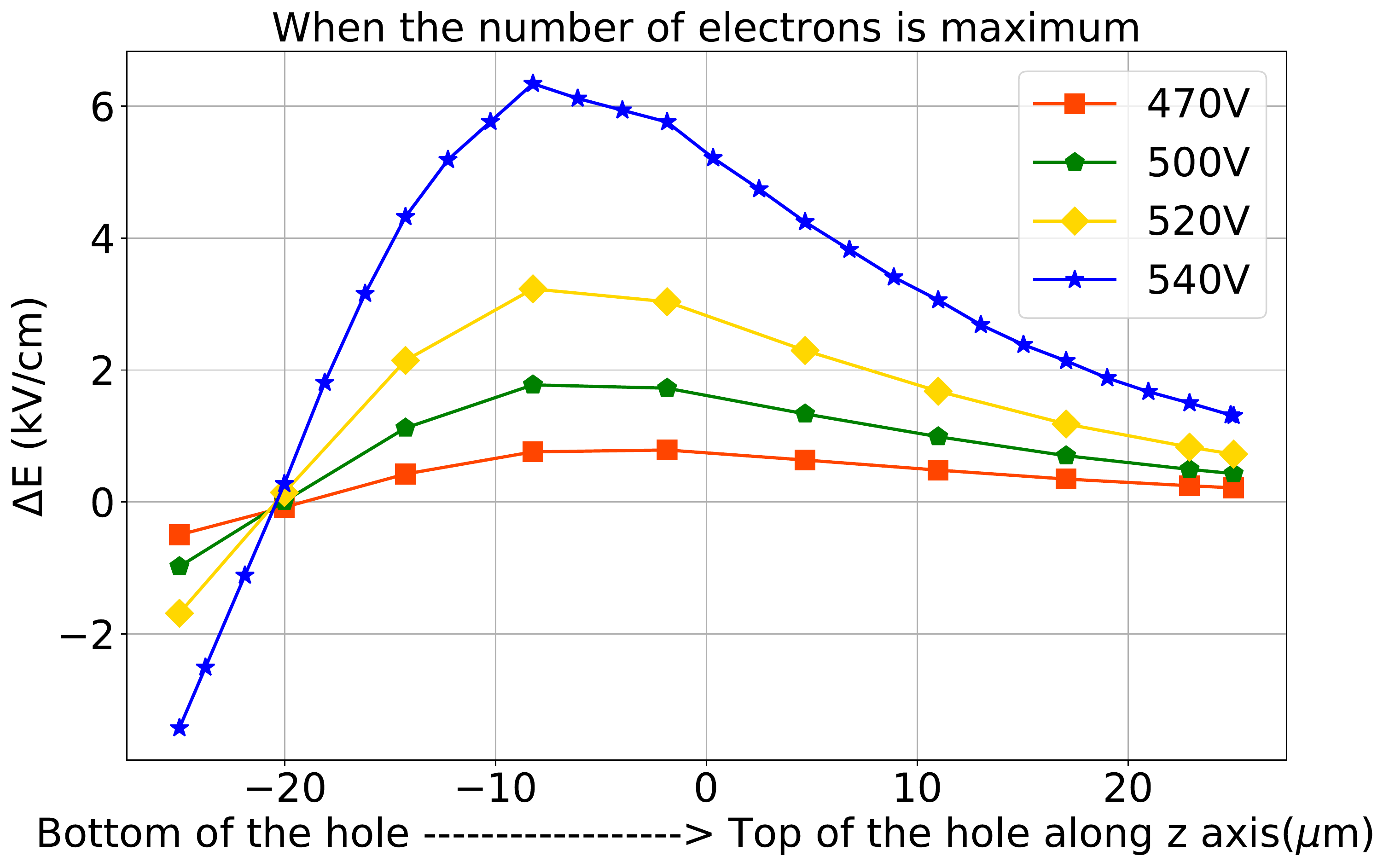}
	\caption{Variation of electric field due to space charge effect with different GEM voltages for cluster case 1}\label{fig6}
\end{figure}

Likewise, space charge effect is found to be increasing with increasing voltages across the GEM as shown in Figure~\ref{fig6}. It shows the variation in electric field due to space charges for case 1 at different GEM voltages. For higher voltages, $\Delta E$ is more and more negative towards the hole bottom which results in the larger decrease in effective gain with respect to the case which does not have any space charge effect.

\section*{Conclusion}

Space charge effect is found to depend significantly on the primary cluster distribution as well as the applied GEM voltages. Sharing of charges among larger number of holes leads to smaller space charge effect. As a result, reduction of gain due to space charge effect is compensated by charge sharing for radially extended clusters and higher cluster heights. Effect of space charge accumulation is found to increase with the increasing applied GEM voltages.






\bibliographystyle{elsarticle-harv}







\section*{Acknowledgement}
    Authors acknowledge their respective institutes for all the support and funding.

\end{document}